\documentclass[11pt,oneside,letterpaper]{article}
\usepackage{amssymb}
\usepackage{amsmath}
\usepackage[dvips]{graphicx}
\usepackage{setspace}
\usepackage{fancyhdr}
\usepackage{xcolor}
\usepackage{ifpdf}
\usepackage{graphicx}
\usepackage{rotating}
\usepackage{comment}
\usepackage{braket}
\usepackage[pdftex,bookmarks,pdfpagelabels,breaklinks,plainpages=false,pdfpagemode=UseNone]{hyperref}
\usepackage{cleveref}

\def\N{\mathcal{N}}

\newcommand{\bi}{\begin{itemize}}
\newcommand{\ei}{\end{itemize}}
\newcommand{\bea}{\begin{eqnarray}}
\newcommand{\eea}{\end{eqnarray}}
\newcommand{\be}{\begin{equation}}
\newcommand{\ee}{\end{equation}}

\newcommand{\Tr}{\text{Tr}}

\numberwithin{equation}{section}
\allowdisplaybreaks  

\begin{document}

\vspace*{2.5cm}
\begin{center}
{ \LARGE \textsc{Grassmann Matrix Quantum Mechanics}}
\\
\vspace*{1.7cm}

Dionysios Anninos$^\triangle$,  Frederik Denef$^{\, \dag,\sharp}$ and Ruben Monten$^{\, \sharp,\dag}$

\vspace*{0.6cm}
\it{\footnotesize $^\triangle$ School of Natural Sciences, Institute for Advanced Study, Princeton, NJ 08540, USA.} \\
\it{\footnotesize $^\dag$ Department of Physics, Columbia University, 538 West 120th Street, New York, New York 10027 }\\
\it{\footnotesize $^\sharp$ Institute for Theoretical Physics, KU Leuven, Belgium}

\vspace*{0.6cm}

\vspace*{0.6cm}


\end{center}
\vspace*{1.5cm}
\begin{abstract}
\noindent

We explore quantum mechanical theories whose fundamental degrees of freedom are rectangular matrices with Grassmann valued matrix elements. We study particular models where the low energy sector can be described in terms of a bosonic Hermitian matrix quantum mechanics. We describe the classical curved phase space that emerges in the low energy sector. The phase space lives on a compact K{\"a}hler manifold parameterized by a complex matrix, of the type discovered some time ago by Berezin. The emergence of a semiclassical bosonic matrix quantum mechanics at low energies requires that the original Grassmann matrices be in the long rectangular limit. We discuss possible holographic interpretations of such matrix models which, by construction, are endowed with a finite dimensional Hilbert space.

\end{abstract}

\newpage
\setcounter{page}{1}
\pagenumbering{arabic}

\setcounter{tocdepth}{2}

\onehalfspacing

\section{Introduction}

Models with matrix like degrees of freedom make numerous appearances throughout physics. Applications range from the study of the spectra of heavy atoms to models of emergent geometry \cite{Kazakov:1986hu,Maldacena:1997re,Banks:1996vh,McGreevy:2003kb,Brezin:1977sv,Klebanov:1991qa}. In this paper we will concern ourselves with a particular class of quantum mechanical models whose degrees of freedom are purely fermionic rectangular matrices $\psi_{Ai}$, with $A = 1, . . . , M$ and $i = 1,...,N$. The matrices transform in the $(M,N)$ bifundamental representation of a $U(M) \times SU(N)$ symmetry group. In a Lagrangian description of the system, transition amplitudes can be expressed as path integrals over Grassmann valued paths  $\psi_{Ai}$. Grassmann matrices naturally appear as the supersymmetric partners of bosonic Hermitian matrices in supersymmetric matrix quantum mechanical theories such as the low energy worldline dynamics of a stack of $N$ D0-branes in type IIA string theory [3, 7] or the Marinari-Parisi matrix model [8]. Our interest is in quantum mechanical models consisting of {\it only} the Grassmann matrices. 

Ordinary integrals over Grassmann matrices were studied extensively in \cite{Makeenko:1993jg,Semenoff:1996vm,Paniak:2000zy}. There, it was shown how the problem of Grassmann matrix integrals at large $N$, $M$ can be expressed as an eigenvalue problem for the composite $N\times N$ matrix $\Phi_{ij} = \sum_A \bar{\psi}_{iA}\psi_{Aj}$, which is effectively bosonic. Unlike bosonic matrices, a Grassmann valued matrix cannot be diagonalized and characterized in terms of eigenvalues. Instead, the authors were able to analyze the model by diagonalizing $\Phi_{ij}$. Certain features of the $\Phi_{ij}$ integral, such as a contribution to the potential of the form $\text{tr} \log \Phi$, were shown to be universal and specifically related to the Grassmann nature of the original problem. Along a similar vein, emergent bosonic matrices from spin systems were considered in  \cite{parisi,Anninos:2014ffa}. 
The models of interest in our work can be viewed as multi-particle quantum mechanical models of fermions which can occupy a finite set of single particle states $|A,i,\alpha\rangle$, labeled by the matrix indices. In particular the Hilbert space is finite dimensional. Fermionic multi-particle models often arise as lattice models in condensed matter physics, where there is typically an assumption about some sort of nearest-neighbour interaction between the fermions reflecting spatial locality. In contrast, the class of models of interest in our paper have no such notion of spatial locality. They are described by actions of the form:
\begin{equation}
S = \int dt \, i \, \sum_{A,\alpha,i} \bar{\psi}^\alpha_{iA} \partial_t \psi^\alpha_{A i} - \text{tr}_{N\times N} \, V\left(\sum_{A,\alpha,\beta} \bar{\psi}^\alpha_{iA} \mathbf{\sigma}_{\alpha\beta} \psi^\beta_{A j }\right)~.
\end{equation}
The potential $V(x)$ is an $N\times N$ matrix valued function. The index $\alpha$ is an spinor index associated to the $d$-dimensional rotation group, but we will focus on the particular case of $d=3$ and take the $\sigma_{\alpha\beta}$ to be the ordinary Pauli matrices. We will also demand that the potential $V(x)$ be $SO(3)$ invariant.\footnote{Part of the reason for choosing an $SO(3)$ index is to mimic the examples of matrix quantum mechanics that appear in holography, where the matrices are labeled by a similar rotational index. We discuss this further in the outlook.}  An example of such a model was studied in \cite{Berenstein:2004hw}. The objects we wish to understand are path integrals over $\{\bar{\psi}^\alpha_{iA}(t),\psi^\alpha_{Ai}(t)\}$ rather than simple integrals. In particular, we study to what extent the Grassmann matrix models at large $N$ and $M$ can be described in terms of a composite bosonic matrix degree of freedom. We then describe several features of the emergent bosonic matrix quantum mechanical systems. We focus on the case where $V(x)$ is quartic in the Grassmann matrices, but the techniques we develop can be used more generally.


As mentioned, our models have a finite dimensional Hilbert space. In this sense they differ from many of the quantum mechanical models studied in the context of holography, such as the D0-brane quantum mechanics or $\mathcal{N}=4$ super Yang-Mills, where the systems have an infinite space of states, even at finite $N$. On the other hand, several proposals have been made throughout the literature suggesting that the holographic dual of a de Sitter universe (or at least its static patch) is indeed a system with a finite dimensional Hilbert space \cite{Banks:2006rx,Parikh:2004wh,Dong:2010pm,Li:2001ky,Volovich:2001rt,Heckman:2011qu}. Our considerations are particularly similar, in spirit, to those of \cite{Banks:2006rx,Parikh:2004wh} where the basic building blocks are also taken to be a large collection of fermionic operators. Part of our motivation is to understand to what extent systems with a finite Hilbert space can give rise to a holographic description with a dual gravitational theory in an appropriate large $N$ type limit. In order for this to be the case, bosonic variables (such as the Hermitean matrices) should emerge from the discrete variables, at least at low energies and in an appropriate large $N$ limit. The models studied in this work serve as toy models where this can be seen explicitly, and we can examine to what extent the bosonic effective degrees of freedom adequately capture the physics and when this description breaks down.

The first part of the paper provides a detailed study for the $N=1$ case, in which the degrees of freedom are organized as vectors. We derive several results regarding the physics of the effective composite degree of freedom $\bar{\psi}^\alpha_A \sigma_{\alpha\beta} \psi^\beta_A$. We show to what extent the theory is described by three bosonic degrees of freedom $\mathbf{x} = (x,y,z)$ transforming as an $SO(3)$ vector. The Euclidean path integral is expressed as a path integral over $\bold{x}$ and a low velocity expansion is developed at large $M$. We study the theories at finite temperature and note a breakdown of the bosonic description at high temperatures. We describe the structure of the emergent classical phase space for the effective bosonic theory, which is the compact K{\"a}hler manifold $\mathbb{C}\mathbb{P}^1$. Some of the results in this section have appeared in several contexts (see for example \cite{abanov,Karchev:2012pg,Aitchison:1986qn}). However, certain aspects of our treatment are novel and furthermore our treatment naturally generalizes to the matrix case. This is studied in the second part of the paper, where now the effective theory becomes that of three bosonic Hermitian $N\times N$ matrices $\mathbf{\Sigma}^a_{ij}$, with $a \in \{ x,y,z \}$. The matrix $\mathbf{\Sigma}^a_{ij}$ transforms in the adjoint of $SU(N)$ and is an $SO(3)$ vector. The matrix analogue of the emergent classical phase space is identified as a compact K{\"a}hler manifold, first introduced by Berezin \cite{Berezin:1978sn}. The K{\"a}hler metric is parameterized by a complex $N\times N$ matrix $Z_{ij}$. We discuss how the $Z_{ij}$ and $Z^\dag_{ij}$ relate to the description of the system in terms of the $\mathbf{\Sigma}^a_{ij}$ as well as the original Grassmann matrices. The volume of the K{\"a}hler metric computes the dimension of the Hilbert space captured by the (quantized) classical phase space. It is shown to precisely match the dimension of the $U(M)$ invariant Hilbert space of the original Grassmann theory. We end with an outlook discussing speculative connections of our models to holography. 


\section{ Vector model}

In this section we discuss a quantum mechanical model in which the degrees of freedom are a vector $\psi^\alpha_A$ of complex Grassmann numbers, with $A = 1,\ldots,M$ and $\alpha=1,2$ a spinor index of $SU(2)$, the double cover of the rotational group $SO(3)$. Our system has a $2^{2M}$ complex-dimensional Hilbert space of states. The purpose of the section is to analyze a simplified version of the matrix model studied in the next section, which however still retains some of the salient features. 

We focus on an action with quartic interactions of the specific form:
\begin{equation}
S = \int dt \, i\, \bar{\psi}_A^\alpha \partial_t \psi^\alpha_A + g \left(\bar{\psi}_A^\alpha \sigma^a_{\alpha\beta} \psi^\beta_{A} \right)\left(\bar{\psi}_B^\gamma \sigma^a_{\gamma\delta} \psi^\delta_{B} \right)~,
\end{equation}
where it is understood that the $A$ and $\alpha$ indices are summed over and the $\sigma^a_{\alpha\beta} = \{ \sigma^x_{\alpha\beta}, \sigma^y_{\alpha\beta}, \sigma^z_{\alpha\beta}\}$ are the three Pauli matrices. The model has an $SU(2)\times U(M)$ global symmetry group. The ($\bar{\psi}^\alpha_A$) $\psi_A^\alpha$ transform in the (anti-)fundamental representation of $U(M)$ and $SU(2)$.

Upon canonical quantization, the non-vanishing anti-commutation relations between the fermionic operators are given by $\{\bar{\psi}_A^\alpha, {\psi}_B^\beta \} = \delta^{\alpha\beta}\delta_{AB}$. The $SU(2)$ generators working on these operators are given by $\mathbf{\hat{J}}^a = \bar{\psi}_A^\alpha \sigma^a_{\alpha\beta} \psi^\beta_{A}/2$. The $U(M)$ generators are given by:
\begin{equation}
\hat{\mathcal{J}}^n = \bar{\psi}^\alpha_A T^n_{AB} \psi^\alpha_B  + c \, \hat{\mathbb{I}} \, \delta^n_0~, \quad\quad n =0,1,\ldots, M^2-1~.
\end{equation}
The $T^n_{AB}$ with $n>0$ are the traceless generators of $SU(M)$ subgroup of $U(M)$, and $T^0_{AB} = \delta_{AB}$ generates the $U(1)$ subgroup of $U(M)$. $c$ is a normal ordering constant that appears as a possible central extension of the $U(1)$. As expected, $[\hat{\mathcal{J}}^n,\hat{\mathbf{J}}^a] = 0$. We take $g>0$ in what follows and measure quantities in units of $g$ so that $g=1$.

\subsection{Spectrum}

The Hamiltonian of the system is proportional to the normal ordered square of the angular momentum operator:
\begin{equation}\label{vecham}
\hat{H} = - : \bar{\psi}^\alpha_A \sigma_{\alpha\beta}^a \psi^\beta_{A} \, \bar{\psi}^\gamma_B \sigma_{\gamma\delta}^a \psi^\delta_{B} : = - 4   :   \mathbf{\hat{J}} \cdot \mathbf{\hat{J}} : = -4 \, \hat{\bold{J}} \cdot \hat{\bold{J}} + 3 \hat{n}~,
\end{equation}
where $\hat{n} \equiv \bar{\psi}_A^\alpha \psi_A^\alpha$, commutes with the $\hat{\bold{J}}^a$. If we view the index $A$ as a lattice site, the system above is describing two-body $SU(2)$ spin-spin interactions of spin-1/2 fermions between all $M$ possible lattice sites, each with equal strength. From (\ref{vecham}), it follows that the the eigenstates $|J,m;n\rangle$ can be labeled by their total angular momentum $J$, their angular momentum $m$ in the $z$-direction and their eigenvalue $n$ with respect to the $\hat{n}$ operator. The energy of $|J,m;n\rangle$ is simply $E = -4J(J+1)+3n$. For $M>1$, the ground states $|g\rangle$ are the $(M+1)$ states in the maximally spinning spin-$M/2$ multiplet, whereas the $J=0$ state with $n=2M$ has maximal energy. We can construct the full Hilbert space by acting with the $\bar{\psi}_{A}^\alpha$ operators on the particular $J=0$ state $|0\rangle$, defined to be the state annihilated by all the $\psi^\alpha_A$. For instance the ground state with maximal spin-$z$ angular momentum is $|M/2,M/2;M \rangle = \prod_A \bar{\psi}^1_A |0\rangle$ and has energy $E_g = -M(M-1)$. 

For each $A$ we have two states with vanishing angular momentum in the $z$-direction, and a spin-$1/2$ doublet. The full Hilbert space can thus be written succinctly as $\mathcal{H} = \left( 0 \oplus 1/2 \oplus 0 \right)^{\otimes M}$. The degeneracies for a given angular momentum in the $z$-direction can be obtained from the partition function:
\begin{equation}
Z[q] = \text{tr} \, q^{\sum_A{\mathbf{J}_A^z}} = \sum_{k=0}^{2M} {2M  \choose k} q^{M/2-k/2}
\end{equation}
From the above partition function, we can also obtain the degeneracies of the multiplets with total spin $J$:
\begin{equation}\label{dj}
d_J = {2M  \choose M + 2J}  - {2M  \choose M + 2(J+1)}~.
\end{equation}
Indeed, there is exactly one state with $m = M/2$, which is part of the maximally spinning (ground state) multiplet. There are $2M$ states with $m = (M-1)/2$, each of which is part of a spin-$(M-1)/2$ multiplet. However, out of the $M(2M-1)$ states with $m = M/2 - 1$, one is already part of the maximally spinning multiplet, leaving $(2M^2 - M - 1)$ spin-$(M-2)/2$ multiplets. Generalizing this argument to all eigenvalues of $\hat{\mathbf{J}}^z$ yields the formula above. As expected, $\sum_J (2J+1) d_J = 2^{2M}$ and $d_{M/2} = 1$. At large $M$, using the Stirling approximation, we find a large degeneracy of $2^{2M}/M$ $J=0$ states. Moreover, for small $J/M$, we can use the approximations: 
\begin{equation}
{2M  \choose M + 2J}  \approx  {2M  \choose M} e^{-4 J^2/M}~, \quad\quad {2M  \choose M + 2(J+1)} \approx  {2M  \choose M} e^{-4 (J+1)^2/M}~. 
\end{equation}
From these we can derive that $d_J$ peaks at $J \approx \sqrt{M/8}$. We show a plot of the degeneracies $d_J$ in figure $\mathbf{1}$. 
\begin{figure}
\begin{center}
{\includegraphics[scale=0.62]{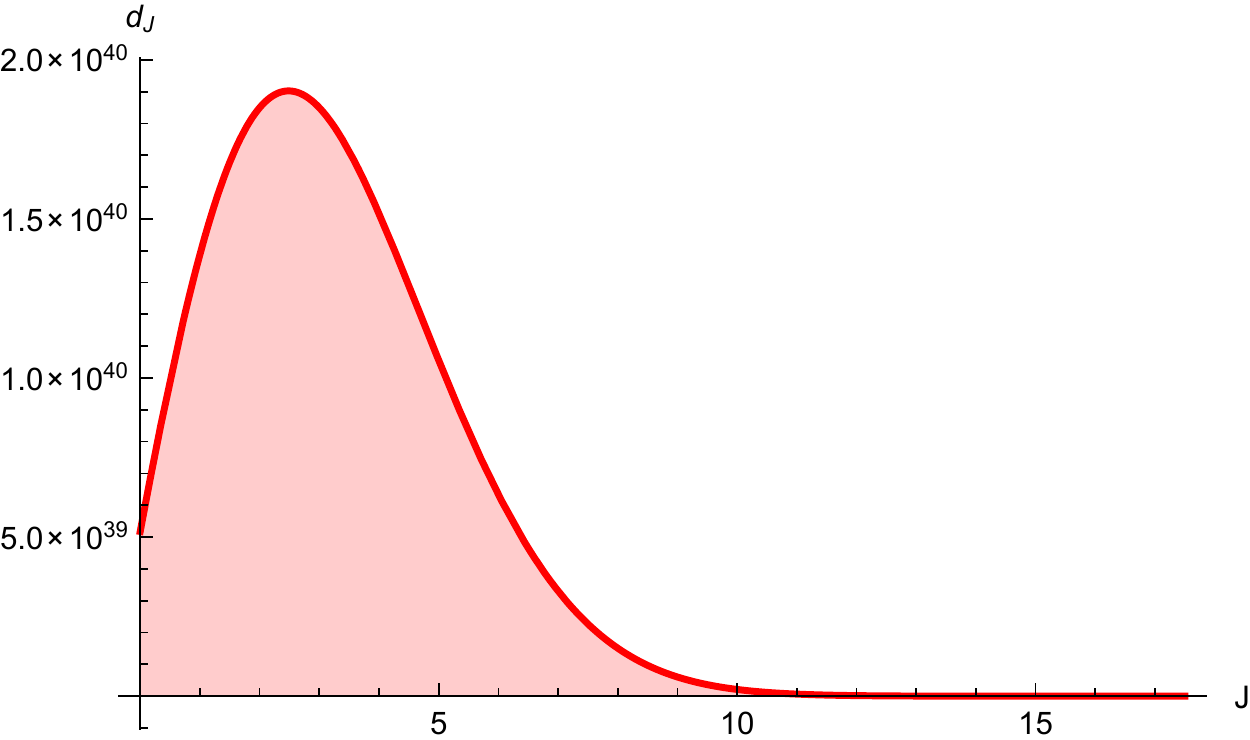}}
\end{center}
\caption{Plot of $d_J$ vs. $J$ for $M=70$. }
\end{figure}

The $d_J$ are the exact degeneracies for the operator $\hat{\tilde{H}} = \left( \hat{H}-3\hat{n} \right)$, with eigenvalues $\tilde{E}_J = -4 J(J+1)$. At large $M$, the $d_J$ are also approximately the degeneracies of $\hat{H}$ for several of its lowest lying states. For example, the energy difference between the ground state with $J=M/2$ and the nearest energy level with $J=(M-1)/2$ is $2M$ to leading order. The $\hat{n}$ operator does not split the energies of the $(M+1)$-fold degenerate states in the ground state multiplet, but it does split the energies of the $2M$ distinct $J=(M-1)/2$ multiplets into two bands of $M$ multiplets separated by an $\mathcal{O}(1)$ amount in energy. Since the energies of both the $J=M/2$ and $J=(M-1)/2$ multiplets are $-M^2$ at large $M$, to leading order in $M$ the $d_J$ are a good approximation of the degeneracies of $\hat{H}$ for the two lowest lying states. More generally, considerations similar to those leading to (\ref{dj}) lead to the formula for the degeneracies of distinct $J$-multiplets with a given $n$:
\begin{equation}
    d_{J,n} = {M \choose \frac{n}{2}+J}{M \choose \frac{n}{2} - J} - {M \choose \frac{n}{2}+J+1}{M \choose \frac{n}{2}-J-1}~,
\end{equation}
where $n = 2J, 2J + 2, \ldots, 2M - 2J$.\footnote{As a simple check, $\sum_n d_{J,n} = d_J$ reproduces \eqref{dj}. Furthermore, $\sum_J d_{J,n}(2J+1) = {2M \choose n}$, where $J = n/2, n/2 - 1, \ldots$ covers positive integer or half-integer values, depending on whether $n$ is even or odd.} When $J \sim 3 M/8$ and below, the energy split among multiplets with the same value of $J$ is large enough to cause overlaps between their energy levels and those of multiplets with different $J$.
For example, the $J=0$ states have energies ranging between $E_0 \in [0,6M]$ which can easily be seen to overlap with the energy levels of the $J=1/2$ states. 

In case we had considered gauging the $U(M)$ symmetry, the spectrum would have changed significantly. For instance, by selecting the normal ordering constant $c = -M$, the only gauge invariant states are the $(M+1)$ maximally spinning ground states. 

\subsection{Effective theory}

We would now like to recast the Euclidean path integral of the theory as a Euclidean path integral of a bosonic (mesonic) variable and understand several features of the model in terms of the bosonic degree of freedom. The Euclidean path integral computes features in the low energy sector the system. For instance, the generating function of vacuum correlation functions is given by:
\begin{equation}
    Z[\xi^\alpha_A,\bar{\xi}^\alpha_A] = \int \mathcal{D}\bar{\psi}^\alpha_A \mathcal{D} \psi^\alpha_A \, e^{-S_E[\bar{\psi},\psi] -\int d\tau \bar{\xi}^\alpha_A \psi^\alpha_A - \int d\tau \bar{\psi}^\alpha_A {\xi}^\alpha_A}~,
\end{equation}
where the Euclidean action $S_E$ is obtained from $-iS$ by a Wick rotation $t = -i \tau$. Upon introducing an auxiliary three-vector $\mathbf{x}$ and integrating out the Grassmann variables, this can be recast as:
\begin{equation}
    Z[\xi^\alpha_A,\bar{\xi}^\alpha_A] = \int \mathcal{D}\mathbf{x} \, \det\left( -\partial_\tau  + \sigma \cdot \mathbf{x} \right)^M \, e^{-\int d\tau \, r^2/4} e^{-\int d\tau\, \xi^\alpha_A \left( -\partial_\tau + \sigma \cdot \mathbf{x} \right)_{\alpha\beta}^{-1} \bar{\xi}^\beta_A}~,
\end{equation}
where $r = |\mathbf{x}|$. From the partition function we can read off the effective action for the $\mathbf{x}$ degree of freedom:
\begin{equation}
S_{eff} = -M \Tr \log \left( -\partial_\tau + \sigma \cdot \mathbf{x} \right) + \int d\tau \, \frac{r^2}{4}~.
\end{equation}
As it stands, the above action is highly non-local in $\tau$. We would like to understand under what conditions this action can approximated by a small velocity expansion.
Generally speaking there is no a priori reason for this to be the case in a quantum system, given that the spectrum is discrete and one cannot continuously change the kinetic energy. However, one may hope that it would be a valid approximation at large $M$. We will see that this is the case.

\subsubsection{Small velocity expansion}

It is useful to diagonalize the $2\times 2$ Hermitian matrix $\mathbf{x} \cdot \sigma$ for each $\tau$. Since the $\sigma$ are traceless, we take some $U \in SU(2)$ such that $ U^\dag \, \sigma \cdot \mathbf{x} \, U = r\, \sigma^z$ for each $\tau$. The $U$ matrix is parameterized by a unit vector $\mathbf{n} = (\sin\theta\cos\phi,\sin\theta\sin\phi,\cos\theta)$. Explicitly:
\begin{equation}
U = \left( \begin{array}{cc}
\cos\frac{\theta}{2} & e^{-i\phi}\sin \frac{\theta}{2} \\
e^{i\phi} \sin \frac{\theta}{2} & - \cos\frac{\theta}{2}  \end{array} \right)~. \label{eq:sphericalCoords}
\end{equation}
It then follows that:
\begin{equation}\label{determinantV}
\det\left( -\partial_\tau  + \sigma \cdot \mathbf{x} \right)^M = e^{M\, \Tr \, \log \left( -\partial_\tau - U^\dag \dot{U} + r \, \sigma^z \right)}~.
\end{equation}
Notice that we can transform the above functional determinant under the time reparameterization symmetry
\begin{equation}\label{timerepvector}
    \tau \to f(\tau)~, \quad r(\tau) \to \dot{f}(\tau) r(f(\tau))~, \quad U(\tau) \to U(f(\tau))~,
\end{equation}
\begin{equation}\label{rhs}
    e^{M\, \Tr \, \log \left( -\partial_\tau - U^\dag \dot{U} + r \, \sigma^z \right)} \to e^{M \Tr \log \dot f} e^{M\Tr \log \left( - \partial_\tau - U^\dag \dot U + r \sigma^z \right)}  \ .
\end{equation}
The first factor on the right-hand side of (\ref{rhs}) is independent of $U$ and $r$ and can be absorbed into the overall normalization of the path integral. The above symmetry can therefore be used to set $r$ to a constant in performing a small velocity expansion of the functional determinant.\footnote{In other words, if we view the symmetries (\ref{timerepvector}) as $(0+1)$-dimensional diffeomorphisms of the worldline, $r(\tau)$ becomes the einbein which can always be gauge fixed to a constant.} It follows from this that no time derivatives will be generated for $r$. 

We expand (\ref{determinantV}) in powers of ${\upsilon}^a \sigma^a = i \, U^\dag \dot{U}$ by expanding the logarithm. The zeroth order term is the effective potential governing $r$. Going to Fourier space, the computation becomes:
\begin{equation}\label{veffvec}
V_{eff} = - M \int \frac{d\omega}{2\pi} \log\left( \omega^2 + r^2 \right) + \frac{r^2}{4} = - M \, r + \frac{r^2}{4}~,
\end{equation}
where we have regulated the $\omega$-integral by differentiating once with respect to $r$ and re-integrating it back while setting the constant of integration to zero. Note that the effective potential is minimized at $r=2 M$ for which $V^{(min)}_{eff} = - M^2$. To leading order in $M$ this agrees with the exact ground state energy of the system $E_g =- M(M+2)$. 

The first order term in the velocity expansion is given by:
\begin{equation}\label{WZW}
    S_{kin}^{(1)} =  -M \int \frac{d\omega}{2\pi} \, \left( -i\omega + r \sigma^z \right)_{\alpha\beta}^{-1}  \, i \sigma^a_{\alpha\beta} \, \tilde{\upsilon}^a({0}) = i\,\frac{M}{2} \int d\tau \left( 1 - \cos\theta \right)\dot{\phi}~,
\end{equation}
where $\tilde{\upsilon}^a(l)$ is the Fourier transform of $\upsilon^a$ at frequency $l$. The linear velocity piece $S_{kin}^{(1)}$ is the phase picked up by a unit charge moving on the surface of a two-sphere, in the presence of a magnetic monopole of strength $M/2$ at the origin. 

Similarly, the quadratic kinetic term is found to be:
\begin{equation}\label{kin2}
S_{kin}^{(2)} = M  \int d\tau \,   \frac{1}{2 r}\,  \left(  {({\upsilon}^x)^2+({\upsilon}^y)^2}\right) = M \int d\tau \,   \frac{1}{8 r}\,  \left( \dot{\theta}^2 + \sin^2\theta \, \dot{\phi}^2  \right)~,
\end{equation}
where in the right-hand side we have expressed the answer in terms of $\mathbf{x}$, but now written in spherical coordinates. The higher order terms can be similarly computed and they contain even powers of time derivatives of the angular variables divided by one less power of $r$.\footnote{In appendix \ref{modvec} we consider a modified vector model where the leading kinetic piece is (\ref{kin2}).}

Denoting the characteristic frequency for some particular motion of $\theta$ and $\phi$ by $\omega_c$, the condition that there is a small derivative expansion is:
\begin{equation}
\omega_c \ll  r~.
\end{equation}
For $r$ near the minimum of the effective potential, we have $\omega_c \ll \, M$. Hence, for large $M$ there is a parametrically large range of frequencies allowing for a small velocity expansion. 


\subsection{Finite temperature}


As was previously noted, the original Grassmann system contains a large number of high energy, i.e. $J=0$, states at large $M$. On the other hand the ground state energy is $E_g = -M(M-1)$. Thus the thermal partition function $Z[\beta] = \Tr \, e^{-\beta \hat{H}}$ at large $\beta$ is dominated by the ground states and goes as:
\begin{equation}
\lim_{\beta\to\infty} Z[\beta] = (M+1) \, e^{M(M-1)\beta}~,
\end{equation}
whereas at small $\beta$ we have simply the dimension of the Hilbert space:
\begin{equation}
\lim_{\beta\to 0} Z[\beta] = 2^{2M}~.
\end{equation}
The transition between these two behaviors occurs at $\beta \sim 1/M$. 
\begin{figure}
\begin{center}
{\includegraphics[scale=0.45]{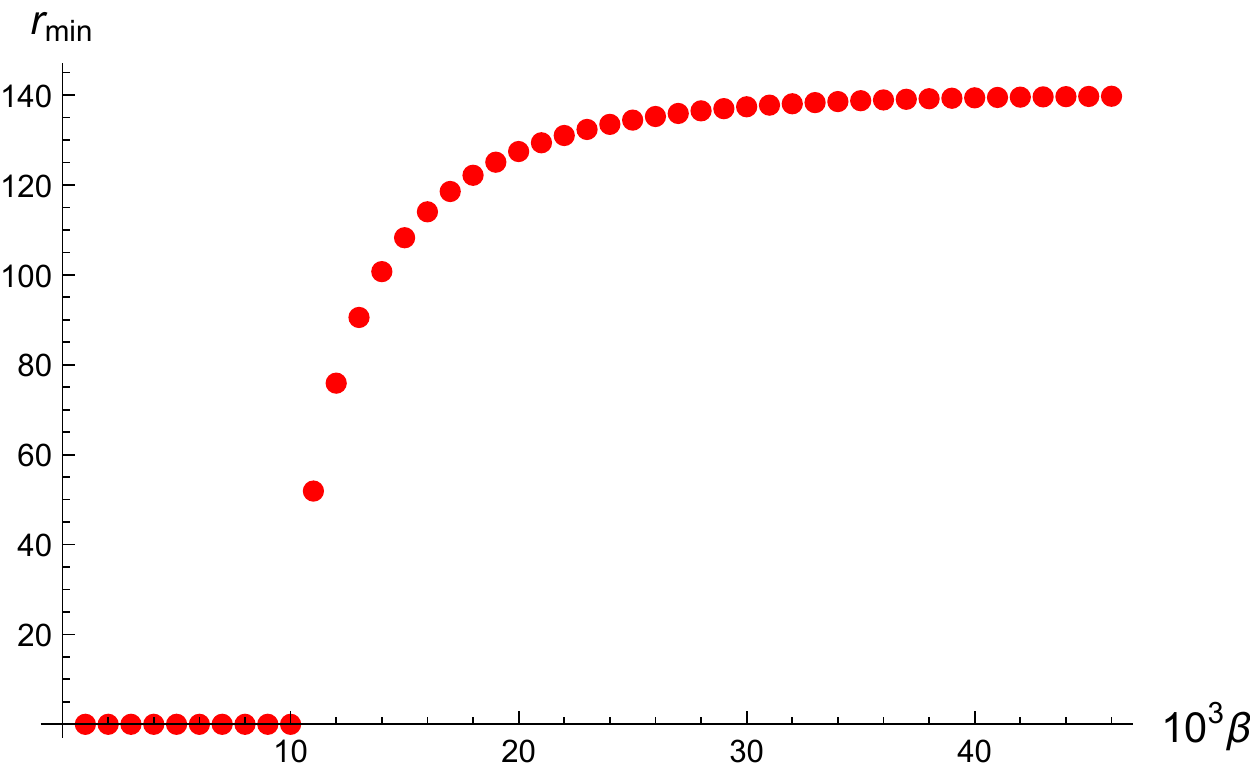}} 
\end{center}
\begin{center}\caption{Plot of value of $r$ minimizing $V_{eff}(\beta)$ vs. $10^3 \times \beta$ for $M=70$. Notice that the value stays close to $2M= 140$ all the way down to $\beta \sim 1/M$.}\end{center}
\end{figure}


We now consider the finite temperature partition function as a Euclidean path integral over $\mathbf{x}$. We must integrate out the Grassmann numbers with anti-periodic boundary conditions along the thermal circle. In analogy to previous calculations, we can compute the thermal effective potential. What changes is that the $\omega$-integrals are replaced by sums over the thermal frequencies $\omega_n = 2\pi(n+1/2)/\beta$ with $n\in \mathbb{Z}$. The thermal effective potential thus becomes:
\begin{equation}
V_{eff}(\beta) = - \frac{M}{\beta} \sum_{n\in \mathbb{Z}} \log \left(  \omega_n^2 + r^2 \right) + \frac{r^2}{4} = - \frac{2 M}{\beta} \log \cosh \frac{r \beta}{2}  + \frac{r^2}{4} ~.
\end{equation}
 As before, the sum has been regulated by differentiating with respect to $r$. 
 
 For large $\beta$, the minimum of $V_{eff}$ is at $r = 2 M$ as for the zero temperature analysis. We can find the critical point for $r$ in a large $\beta$ expansion. To first order:
 \begin{equation}
 r = 2 M \left(1 - 2 e^{-2 M \beta} + \ldots \right)~.
 \end{equation}
From this we see the tendency of $r$ to decrease upon increasing the temperature. At small $\beta$, we can Taylor expand:
\begin{equation}
V_{eff}(\beta) = \frac{r^2}{4}  - \frac{\beta}{4} \, {M \, r^2}  + \mathcal{O}(\beta^2)~.
\end{equation}
We see that for $\beta \lesssim 1/M$ the thermal potential is minimized at $r=0$. In figure  $\mathbf{2}$ we show a plot for the values of $r$ minimizing $V_{eff}(\beta)$ as we vary $\beta$.

When $r$ is near zero, we can no longer assume that the kinetic contributions are small and thus our analysis breaks down. This as an indication that the high temperature phase does not have a reliable small velocity description in terms of $\mathbf{x}$. Instead, the correct description requires taking into account the full set of Grassmann degrees of freedom. 

\subsection{Bloch coherent state path integral}

So far we have introduced the variable $\mathbf{x}$ as a convenient integration variable to capture correlations in the vacuum state and thermal properties. Here we would like to point out that in a fixed large angular momentum sector, there is some more significance to $\mathbf{x}$. 

Following Bloch, we define a collection of coherent states built from the state $\ket{v}$, which has the lowest angular momentum in the $z$-direction and hence is also a minimal energy state. In other words $|v\rangle = \prod_A \bar{\psi}_A^2 |0\rangle$. We can act on $\ket{v}$ with the spin raising operator $\mathbf{\hat{J}}^+ = \mathbf{\hat{J}}^x + i \, \mathbf{\hat{J}}^y$ to generate states in the maximally spinning multiplet,
\begin{equation}
|\bar{z}\rangle  = \frac{1}{\left( 1 + z\bar{z} \right)^{M/2}} \, e^{\bar{z} \, \mathbf{\hat{J}}^+ } | v\rangle~, \quad\quad z \in \mathbb{C} \ .
\end{equation} 
These states are not orthogonal, but they constitute an over-complete basis of the Hilbert space of the maximally spinning multiplet,
\begin{align}
    \braket{w | \bar z} &= \frac{(1 + w \bar z)^M}{(1 + w \bar w)^{M/2} (1 + z \bar z)^{M/2}}   \ , &   \int{d^2z\ \frac{M+1}{\pi (1 + z \bar z)^2} \ket{\bar z} \bra{z}} &= \mathbb{I} \ .
\end{align}
The purpose of these states is to describe, with minimal uncertainty, points on the $S^2$ of spin directions. Indeed, the angular momentum expectation value defines a point on $S^2$ -- through the stereographic projection -- with decreasing uncertainty in the large $M$ limit
\begin{align}
    \mathbf{J}^a \equiv  \braket{z |\mathbf{ \hat J}^a | \bar z} &= \frac{M}{2 (1+|z|^2)} \left( z + \bar z, i (\bar z - z), |z|^2 - 1 \right)   \ , \label{eq:spinExpectation} \\
    \frac{\braket{z | (\mathbf{\hat J}^a - \mathbf{J}^a)^2 | \bar z}}{\braket{z | \mathbf{\hat J}^a | \bar z}^2} &= \frac2M    \ . \nonumber
\end{align}
One may ask about transition amplitude between two such states: $\langle z_N | e^{-i T \hat{H}} | \bar{z}_0 \rangle$ for some given Hamiltonian $\hat{H}$ built out of the $\hat{\bold{J}}^a$. The result is \cite{haldane1,haldane2}:
\begin{equation}\label{blochpi}
\langle z_N | e^{- i T \hat{H}} | \bar{z}_0 \rangle = \int \mathcal{D}z\mathcal{D}\bar{z} \, \frac{\left(M+1\right)}{\pi(1+z \bar{z})^2} \, e^{i S(z,\bar{z})}~, 
\end{equation}
with 
\begin{equation}\label{blochkin}
S = i \frac{M}{2} \int dt \, \frac{\left(   z \dot{\bar{z}}- \dot{z} \bar{z} \right) }{1+ z\bar{z} } - \int dt \, H(z,\bar{z})~, 
\end{equation}
where $H(z,\bar{z}) \equiv \langle z | \hat{H} | \bar{z} \rangle$. The boundary conditions are $z(T)=z_N$ and $\bar{z}(0) = z_0$.  For our particular choice of Hamiltonian, $H(z,\bar{z}) = -M(M-1)$.
Given the first order form of the action (\ref{blochkin}) appearing in the path integral (\ref{blochpi}), the complex variable $z$ can be viewed as a complex coordinate parameterizing a two-dimensional phase space. From the linear velocity piece in (\ref{blochkin}) we note that the phase space is curved and compact, with K{\"a}hler metric:
\begin{equation}\label{blochsphere}
ds^2 = 2M \, \frac{dz d\bar{z}}{(1+z\bar{z})^2}~.
\end{equation}
This is the Fubini-Study metric on $\mathbb{C}\mathbb{P}^1 \cong S^2$, and we occasionally refer to it as the Bloch sphere. The symplectic form is given by the K{\"a}hler form and the large $M$ limit plays the role of the small Planck constant limit. Time evolution of a function $A(z,\bar{z})$ in the emergent classical phase space is governed by the Poisson bracket, i.e. $\dot{A}(z,\bar{z}) =  \{ A(z,\bar{z}) , H(z,\bar{z}) \}_{p.b.} = i\, M^{-1} (1+z\bar{z})^{2} \left( \partial_{\bar{z}} H \partial_{{z}}  A - \partial_{\bar{z}} A \partial_{{z}} H \right)$. The $SU(2)$ symmetry of the original Grassmann model acts on $z$ as:
\begin{equation}\label{ztransform}
z \to {(\alpha z +\beta)}{(\gamma z+\delta)^{-1}}~, \quad\quad \left( \begin{array}{cc}
\alpha & \beta \\
\gamma & \delta \end{array} \right) \cdot
\left( \begin{array}{cc}
\alpha & \beta \\
\gamma & \delta \end{array} \right)^{\dag} = \mathbb{I}_{2\times 2}
~.
\end{equation}
Since the classical phase space has finite volume, we recover the fact that the underlying system has a finite number of ground states.
The complex coordinate $(z, \bar z)$ can be related to the spherical coordinates introduced in \eqref{eq:sphericalCoords} by identifying the expectation value \eqref{eq:spinExpectation} with the bosonic variable $\mathbf{x}$ introduced in the previous section. The stereographic projection then gives $z = e^{i\phi} \cot{\theta/2}$. With this identification, the linear velocity term in (\ref{blochkin}) becomes precisely the one found in (\ref{WZW}). Thus, we see that certain transition amplitudes are captured by a real time path integral between different points localized on an $S^2$. This allows for physical interpretation of the $(\theta,\phi)$ coordinates as real time degrees of freedom, rather than merely integration variables.
\begin{figure}
\begin{center}
{\includegraphics[scale=0.52]{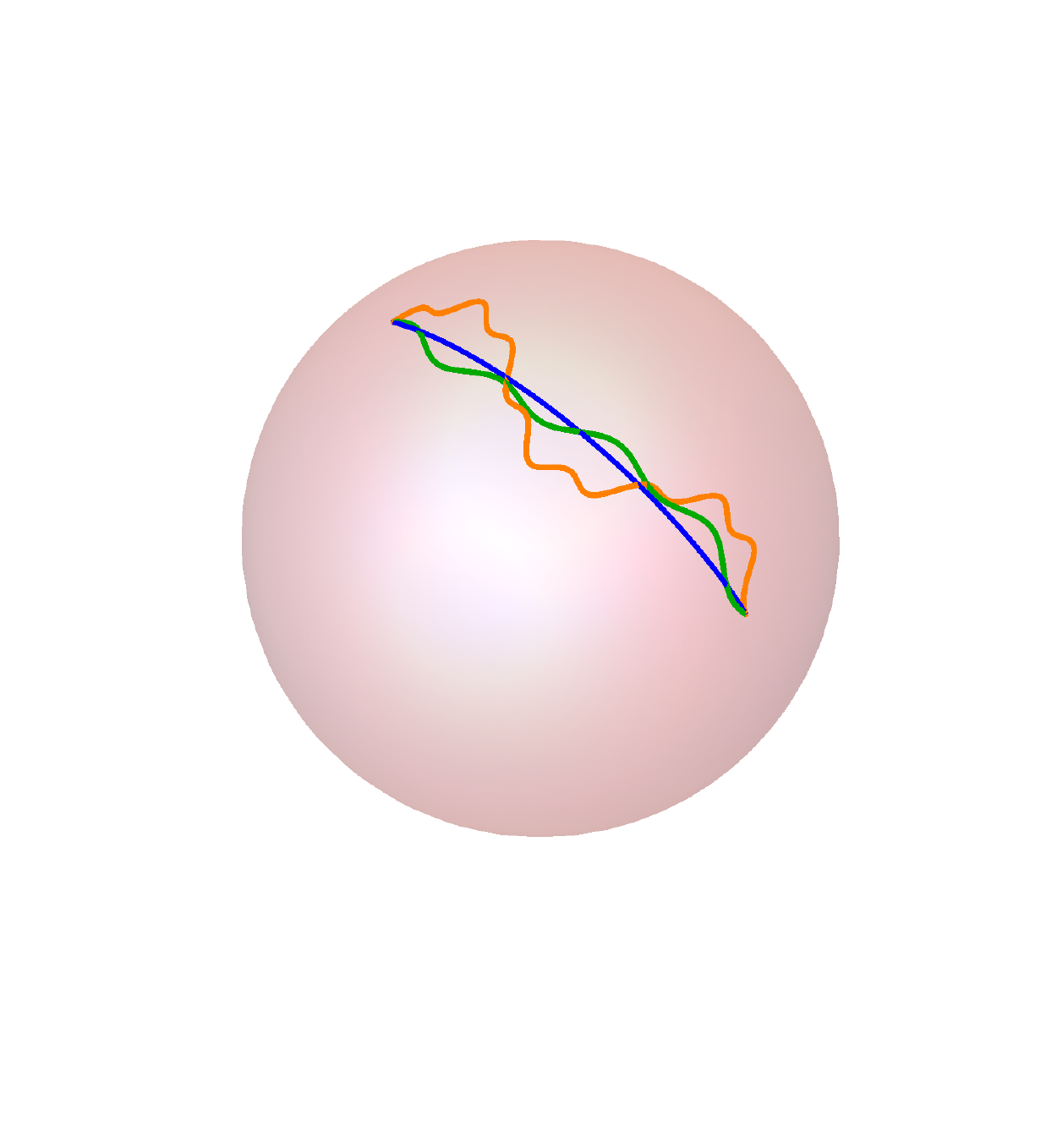}}
\end{center}
\caption{Schematic plot of classical and nearby trajectories on the Bloch sphere for some $H(z,\bar{z})$, contributing to the path integral (\ref{blochkin}). At large $M$ the classical trajectory dominates.}
\end{figure}

We can quantize this low energy effective theory to leading order in the velocity expansion. This becomes the quantum mechanics of an electrically charged particle with unit charge. Its motion is confined to a unit sphere in the presence of a magnetic monopole of strength $M/2$ at the origin. Thus, to leading order in $M$ the ground states are given by the $M$ lowest Landau levels, each with energy $E_g = - M^2$ for our choice of Hamiltonian. Due to the Dirac quantization condition, we recover that $M$ must be an integer. 
\newline\newline
We have seen how certain low energy features in the original Grassmann theory are described in the language of the effective bosonic degree of freedom $\mathbf{x}$. Instead of maximally spinning states built out of anti-commuting creation operators, we have lowest Landau levels of a charged particle. The energies (at least in the the low energy regime) are registered by the absolute value of $\mathbf{x}$. We have observed the breakdown of the bosonic effective theory at high temperatures. Certain features were particular to our model. But others such as the presence of linear velocity terms and the absence of a kinetic term for $r$ may be general features of a larger class of models. At this point we proceed to generalize these observations to the case where we have a matrix worth of Grassmann degrees of freedom.

\section{Matrix model}

The goal of this section is to analyze a matrix version of the vector model studied above. Given that the model is more complicated, we will not be able to attain as explicit a description, however we will uncover and generalize several of the features found in the vector model.

\subsection{Action and Hamiltonian}

Our degrees of freedom are now $2MN$ complex rectangular Grassmann matrices, $\bar{\psi}_{iA}^\alpha$ and  $\psi^\alpha_{Ai}$, with $A = 1,\ldots,M$ and $i=1,\ldots,N$. As before,  $\alpha$ is an $SU(2)$ spinor index. The dimension of the Hilbert space now becomes $2^{2N M}$. The Grassmann elements obey the anti-commutation relations $\{ \psi^\alpha_{Ai}, \bar{\psi}^{\beta}_{jB} \} = \delta^{\alpha\beta}\delta_{ij}\delta_{AB}$. 

We will focus on the following action:\footnote{We have and will continue to suppress the $SU(2)$ spinor index in $\psi^\alpha_{Ai}$ to avoid cluttering of indices.}
\begin{equation}
S = \int d t \, i \, \bar{\psi}_{iA}  {\partial}_t \psi_{Ai}  + g \, (\bar{\psi}_{iA} \mathbf{\sigma}^a \psi_{Aj})  (\bar{\psi}_{jB} \mathbf{\sigma}^a \psi_{Bi})~.
\end{equation}
When $N=1$, the above action reduces to the one analyzed in the previous section. The model exhibits a $U(M)\times SU(N) \times SU(2)$ global symmetry. The $SU(2)$ acts by simultaneously rotating all the Grassmann elements. The capitalized index of ($\bar{\psi}_{iA}^\alpha$) $\psi_{Ai}^\alpha$ transforms in the (anti-)fundamental representation of $U(M)$ whereas the lower case index transforms in the (anti-)fundamental of $SU(N)$. 

The Hamiltonian of the model is given by:
\begin{equation}\label{matrixH}
\hat{H} = - g \sum_{i,j,A,B} : \bar{\psi}_{iA} \mathbf{\sigma} \psi_{Aj} \bar{\psi}_{jB} \mathbf{\sigma} \psi_{Bi} :
\end{equation}
If we view the $A$ index as a lattice site, our system describes $SU(2)$ spin-spin interactions of the spin-1/2 fermions. But now the fermions are labeled by an additional quantum number, the color index $i=1,2,\ldots,N$, which can be exchanged through the interaction. Since interactions between all lattice sites have the same strength, the model exhibits no notion of spatial locality. 

We will analyze $g>0$ and from now on choose units setting $g=1$. Unlike the vector case previously studied, the combinatorial problem of finding the exact spectrum of $\hat{H}$ seems to be rather difficult and we have not solved it. Instead, we will try to extract information about the low energy sector of the theory by going to an effective description in terms of bosonic matrices. Before doing so, we will establish some further properties about the operator algebra. 



\subsubsection{$U(2N)$ operator algebra}
The analogues of the spin operators $\mathbf{\hat{J}}^a =\sum_A \bar{\psi}_A \mathbf{\sigma}^a \psi_A/2$ studied in the previous section are the $U(M)$ invariant $N\times N$ spin matrix operators: $\mathbf{\hat{S}}^a_{ij} = \sum_A (\bar{\psi}_{iA} \mathbf{\sigma}^a \psi_{Aj})/2$. These operators transform as vectors in the three-dimensional real representation of $SU(2)$, as well as in the adjoint of the $SU(N)$. Introducing an additional operator $\mathbf{\hat{S}}^0_{ij} = \sum_A (\bar{\psi}_{iA} \mathbf{\sigma}^0 \psi_{Aj})/2$, with $\sigma^0$ the $2\times 2$ identity matrix, we have the following closed operator algebra:
\begin{eqnarray}
[\mathbf{\hat{S}}_{ij}^a,\mathbf{\hat{S}}_{kl}^b] &=&  \frac{1}{2} \delta^{ab} \left( \delta_{kj} \mathbf{\hat{S}}_{il}^{0} - \delta_{il} \mathbf{\hat{S}}^0_{kj} \right) + \frac{i}{2} \epsilon^{abc} \left( \delta_{kj} \mathbf{\hat{S}}_{il}^{a} + \delta_{il} \mathbf{\hat{S}}^b_{kj} \right)~,  \\
\, [\mathbf{\hat{S}}_{ij}^0,\mathbf{\hat{S}}_{kl}^a] &=& \frac{1}{2} \left( \delta_{kj} \mathbf{\hat{S}}_{il}^{a} - \delta_{il} \mathbf{\hat{S}}^a_{kj} \right)~, \\
\, [\mathbf{\hat{S}}_{ij}^0,\mathbf{\hat{S}}_{kl}^0] &=& \frac{1}{2} \left( \delta_{kj} \mathbf{\hat{S}}_{il}^{0} - \delta_{il} \mathbf{\hat{S}}^0_{kj} \right)~.
\end{eqnarray}
The $N$ diagonal components of the $\mathbf{\hat{S}}_{ij}^a$ generate $N$ copies of the usual $su(2)$ algebra. The above operators can be arranged in a $2N \times 2N$ Hermitian matrix $\sigma_{\alpha \beta}^\mu \otimes \mathbf{\hat{S}}_{ij}^\mu$ (with $\mu = \{ 0,x,y,z \}$ summed over) and hence they generate a $u(2N)$ algebra. They act as $\psi^\alpha_{Ai} \to \psi^\alpha_{Ai}  \, G^{\alpha\beta}_{ij}$ and $\bar{\psi}^\alpha_{iA} \to (G^{\alpha\beta}_{ij})^{-1} \, \bar{\psi}^\beta_{jB}$ with $G^{\alpha\beta}_{ij} = e^{i\lambda^{\alpha\beta}_{ij}} \in U(2N)$ and $\lambda^{\alpha\beta}_{ij} = \lambda_{ij}^\mu \sigma^\mu_{\alpha\beta}$ the elements of a $2N\times 2N$ Hermitian matrix. 

The $U(2N)$ symmetry manifestly commutes with the $U(M)$ group and preserves the anti-commutation relations between the $\psi^\alpha_{Ai}$ and $\bar{\psi}^\alpha_{iA}$. Our Hamiltonian (\ref{matrixH}) does not commute with the full $U(2N)$ but rather the $U(N)$ diagonal subgroup generated by the $\mathbf{\hat{S}}^0_{ij}$. When $N=1$, the $U(2N)$ algebra becomes nothing more than the global $SU(2)$ symmetry of the vector model, which not only commutes with the $U(M)$ global symmetry but also with the Hamiltonian. 


%
%
%
\subsection{Effective theory}\label{gsmatrix}

We introduce three $N\times N$ Hermitian bosonic matrices $\mathbf{\Sigma}_{ij}^a=(\mathbf{\Sigma}^x_{ij},\mathbf{\Sigma}^y_{ij},\mathbf{\Sigma}^z_{ij})$. In analogy with the vector case, we introduce them as auxiliary variables which are given on-shell by $\mathbf{\Sigma}^a_{ij} = 2 \,  \mathbf{\hat{S}}_{ij}^a$. Upon integrating out the $\psi_{Ai}^\alpha$, the generating function of vacuum correlations of $\psi$ and $\bar{\psi}$ can be expressed as a Euclidean path integral over the $\mathbf{\Sigma}_{ij}$:
\begin{equation}\label{partitionM}
Z[\xi_{Ai}^\alpha,\bar{\xi}_{iA}^\alpha] = \int \mathcal{D} \mathbf{\Sigma} \, e^{M \text{Tr} \log \left( -\partial_\tau  + \mathbf R \right) - \frac{1}{4} \text{tr}  \, \int d\tau \, \mathbf{\Sigma} \cdot \mathbf{\Sigma}} \, e^{\int d\tau \, \bar{\xi}_{iA}^\alpha \left( -\partial_\tau + \mathbf R \right)_{ij,\alpha\beta}^{-1} \xi_{A j}^\beta}~.
\end{equation}
We have defined $\mathbf R \equiv \mathbf{\Sigma}^x \otimes \sigma^x +  \mathbf{\Sigma}^y \otimes \sigma^y +  \mathbf{\Sigma}^z \otimes \sigma^z$. We also denote the full functional trace by $`\text{Tr}$' and reserve the $`\text{tr}$' symbol for the ordinary matrix trace. It follows from this definition that $\text{tr} \, \mathbf R = 0$.
The global $SU(N)$ symmetry acts as $\mathbf{\Sigma} \to U \mathbf{\Sigma} U^\dag$. Also, $\mathbf{\Sigma}$ transforms as in the three-dimensional (vector) representation of the global $SU(2)$ symmetry group. We can also write down the generating function for vacuum correlations of the composite spin-matrix operator $\mathbf{\hat{S}}^a_{ij}$. These are computed by the correlation functions of $\mathbf{\Sigma}_{ij}$ itself:
\begin{equation}
Z[\mathbf{J}_{ij}^a] = \int \mathcal{D} \mathbf{\Sigma} \, e^{M \text{Tr} \log \left( -\partial_\tau  + \mathbf R  \right) - \frac{1}{4} \text{tr}  \, \int d\tau \, \mathbf{\Sigma} \cdot \mathbf{\Sigma}} \, e^{\frac{1}{4}\text{tr} \, \int d\tau \mathbf{J} \cdot \mathbf{\Sigma} - \frac{1}{16} \text{tr} \, \int d\tau \mathbf{J} \cdot \mathbf{J}}~,
\end{equation}
where $\mathbf{J}^a_{ij}$ are sources for the $\mathbf{\hat{S}}^a_{ij}$. It is worth noting that, unlike the $N=1$ case, the $\mathbf{\hat{S}}^a_{ij}$ no longer commute with the Hamiltonian and thus non-trivial time correlations amongst them may exist. 

We now proceed to study the validity and properties of the `small velocity' expansion of $\det \left( -\partial_\tau  + \mathbf R \right) = \exp \left[ \text{Tr} \log \left( -\partial_\tau + \mathbf R \right) \right]$. Since $\mathbf R$ is a $2N\times 2N$ Hermitian matrix, we can diagonalize it as $U^\dag \mathbf R U = \lambda$ with $\lambda = \text{diag} \left[\lambda_1,\ldots,\lambda_{2N}\right]~$, $U \in U(2N)$ and $\lambda_n \in \mathbb{R}$. Note that due to the tracelessness of $\mathbf R$, not all $\lambda_n$ can have the same sign. Similar to the $N=1$ case, in the diagonal $\mathbf R$ frame, we can write the functional determinant as:
\begin{equation}\label{fdetu2n}
\text{Tr} \log \left(-\partial_\tau + \mathbf R  \right)  = \text{Tr} \log \left(-\partial_\tau - U^\dag \dot{U} + \lambda  \right)~.
\end{equation}
With the above expression we can again use the time reparameterization symmetry
\begin{equation}\label{timereps}
\tau \to f(\tau)~, \quad\quad \lambda_n(\tau) \to f'(\tau) \lambda_n(f(\tau))~, \quad U(\tau) \to U(f(\tau))~,
\end{equation}
to see that the effective action will be independent of $\dot{\lambda}_n$, analogous to how the vector model is independent of $\dot{r}$. Using the propagator:
\begin{equation}
G(\omega) = \text{diag} \left[\left( -i\omega+ \lambda_1\right)^{-1},\ldots,\left( -i\omega+ \lambda_{2N} \right)^{-1} \right]~,
\end{equation}
we can expand the logarithm in powers of the Hermitian matrix $\upsilon = i U^\dag \dot{U}$. Each term in the expansion will be endowed with a $U(2N)$ symmetry taking $U^\dag \dot{U} \to \Lambda^\dag \, \left( U^\dag \dot{U} \right) \,  \Lambda$ and $\lambda \to \Lambda^\dag \, \lambda \, \Lambda$ with $\Lambda \in U(2N)$.

The linear velocity contribution to the effective action is:
\begin{equation}
    S^{(1)}_{kin} = -i M \, \text{tr} \, \int \frac{d\omega}{2\pi} G(\omega) \, \tilde{\upsilon}({0}) = -i \, \frac{M}{2} \sum_{m} \text{sgn} (\lambda_m)\, \int d\tau \,  \left[i \, U^\dag \dot{U} \right]_{mm}~.
\end{equation}
The $\tilde{\upsilon}({l})$ is the Fourier transform of $\upsilon$ at frequency $l$. To define the above $\omega$-integral we have put a cutoff at large $\omega$, performed the exact integration and then taken the large cutoff limit. The kinetic piece containing two time derivatives in $U(\tau)$ is given by:
\begin{align}\label{kin2matrix}
    S^{(2)}_{kin} &= - \frac{M}{2} \, \text{tr} \, \int \frac{d\omega \, dl}{(2\pi)^2} \, G(\omega) \,  \tilde{\upsilon}({l})  \, G({\omega}) \, \tilde{\upsilon}({-l}) \nonumber \\
                  &=  \frac{M}{2} \, \sum_{n,m} \, \int d\tau \, \left[ i\, U^\dag \dot{U} \right]_{nm} \Lambda_{mn} \left[ i\, U^\dag \dot{U} \right]_{mn}~,
\end{align}
with $\Lambda_{mn} = 1/|\lambda_m-\lambda_n|$ and the sum running only over the pairs $(n,m)$ for which $\lambda_n$ and $\lambda_m$ have opposite signs. The reason why only pairs of $\lambda_m$ with opposite sign appear in the sum is that the integral appearing in (\ref{kin2matrix}):
\begin{equation}
\mathcal{I}_{mn} =  \int \frac{d\omega}{2\pi} \frac{1}{\left(-i\omega + \lambda_m\right)} \, \frac{1}{\left(-i\omega + \lambda_n\right)} 
\end{equation}
vanishes whenever $\lambda_n$ and $\lambda_m$ have the same sign. It is interesting to note that the effective kinetic piece of the theory, and hence what we mean by the dynamical content, depends on the particular distribution of eigenvalues $\lambda_n$.

Having obtained expressions for the first few velocity dependent terms in the effective action, we can estimate when the low velocity expansion is valid. Denoting the characteristic frequency for some motion as $\omega_c$, then in order for $S^{(1)}_{kin}$ to be large compared to $S^{(2)}_{kin}$ one requires:
\begin{equation}
\omega_c \ll \frac{\lambda_n}{N}~.
\end{equation}
The factor of $N$ stems from the fact that $S^{(2)}_{kin}$ has an additional matrix index to be summed over that was not present in the vector model previously studied. 
In what follows we will see that the effective potential is minimized for $\lambda_m \sim M$. Thus, in the limit $M \gg N$, we can have a large range of allowed $\omega_c$ (in units where $g=1$). If instead $M$ does not scale with $N$ and we take the large $N$ limit, the window of allowed $\omega_c$ shrinks to zero. 

Since the global symmetry group of the theory, for our choice of Hamiltonian, is {\it not} the full $U(2N)$, the situation is not as simple as the $N=1$ case. For instance, the $\mathbf{\Sigma}$ measure in the path integral is not $U(2N)$ invariant. Moreover, it is in general complicated to quantify how the $\mathbf{\Sigma}$ matrices are encoded in the $\lambda_n$ eigenvalues and $U$ matrices. In what follows we express several parts of the effective action directly in terms of the $\mathbf{\Sigma}$. 



\subsubsection{Effective potential}\label{matrixveff}

We would now like to focus on the effective potential $V_{eff}$ for $\mathbf{\Sigma}$. In order to compute this we can take $\mathbf{\Sigma}$ to be time independent. $V_{eff}$ must respect the $SU(N)\times SU(2)$ symmetries. For instance it can contain a piece which is the trace of a function of the $SU(2)$ invariant matrix $\mathbf{\Sigma} \cdot \mathbf{\Sigma}$. Moreover, when the $\mathbf{\Sigma}$ are diagonal (or when they all commute with each other), it must reproduce $N$ copies of the potential (\ref{veffvec}) we found in the vector model. Finally, the piece of $V_{eff}$ originating from the functional determinant must scale linearly in $\mathbf{\Sigma}$. We can write a general expression by noting that:
\begin{equation}
{\det}_{2N\times 2N} \left(- i \omega + \mathbf R \right) = \prod_{n=1}^{2N} \left( - i \omega +  \lambda_n  \right)~, 
\end{equation}
is the characteristic polynomial for matrix $\mathbf R$ with eigenvalues $\lambda_n$. We must also take the product over all $\omega$, a procedure which must be regulated. For each $\lambda_n$, we can express the product over the $\omega$ as the exponential of an integral over the logarithm:
\begin{equation}
\frac{1}{2} \, \int \frac{d\omega}{2\pi} \log \left(  \omega^2 + \lambda^2_n \right) = \frac{|\lambda_n|}{2}~.
\end{equation}
To define the above integral,\footnote{One may be concerned about the discontinuity of the first derivative at $\lambda_n = 0$. However, the expression agrees with what we expect of the determinant $\prod_\omega (1+  \lambda_n^2/\omega^2)$. Namely, it should equal one when $\lambda_n = 0$, it should be symmetric under $\lambda_n \to - \lambda_n$ and have an exponent linear in $\lambda_n$. Moreover, one can check that at any non-zero temperature $T$ for which $\omega \to 2\pi T (n+1/2)$ with $n \in \mathbb{Z}$, the kink at $\lambda_n = 0$ smoothens out.} we have subtracted the integral of $\log (\omega^2)$. Putting things together:
\begin{equation}\label{veffsigma}
V_{eff} = - \frac{M}{2} \sum_{n=1}^{2N}  |\lambda_n| + \frac{1}{4}  \text{tr} \, \mathbf{\Sigma} \cdot \mathbf{\Sigma} = -\frac{M}{2} \text{tr} \sqrt{ \mathbf R^2 } + \frac{1}{4}  \text{tr} \, \mathbf{\Sigma} \cdot \mathbf{\Sigma}~.
\end{equation}
As expected, $V_{eff}$ is invariant under both the $SU(N)$ and $SU(2)$ global symmetries. It is instructive to write the $2N\times 2N$ matrix $\mathbf R^2$ explicitly:
\begin{equation}
\mathbf R^2 = 
\left( \begin{array}{ccc}
\mathbf{\Sigma}\cdot \mathbf{\Sigma} - i [\mathbf{\Sigma}^x,\mathbf{\Sigma}^y] & [\mathbf{\Sigma}^z,\mathbf{\Sigma}^x+i\mathbf{\Sigma}^y] \\
 -[\mathbf{\Sigma}^z,\mathbf{\Sigma}^x-i\mathbf{\Sigma}^y]  & \mathbf{\Sigma}\cdot \mathbf{\Sigma} + i [\mathbf{\Sigma}^x,\mathbf{\Sigma}^y]  \end{array} \right)~.
\end{equation}
From the above expression, it immediately follows that $\text{tr} \mathbf R^2 = 2 \, \text{tr} \, \mathbf{\Sigma} \cdot \mathbf{\Sigma}$. However, this does {\it not} imply that $\text{tr} \sqrt{ \mathbf R^2 } = 2 \,  \text{tr} \sqrt{ \mathbf{\Sigma} \cdot \mathbf{\Sigma} }$ unless all the $\mathbf{\Sigma}$ commute amongst each other. Thus, we see  how the commutator interaction enters the potential. If it happens that the $\mathbf{\Sigma}$ are almost commuting, we can perform a matrix Taylor expansion of $\text{tr} \sqrt{ \mathbf R^2}$, which to leading order gives:
\begin{equation}\label{matrixTaylor}
    -\frac M2 {\text{tr}} \sqrt{ \mathbf R^2} \approx -M \text{tr} \sqrt{ \mathbf{\Sigma} \cdot \mathbf{\Sigma} } + \frac{M}{16} \text{tr} (\mathbf{\Sigma} \cdot \mathbf{\Sigma})^{-1/2} i [\mathbf{\Sigma}^a,\mathbf{\Sigma}^b] (\mathbf{\Sigma} \cdot \mathbf{\Sigma})^{-1} i [\mathbf{\Sigma}^a,\mathbf{\Sigma}^b]  +  \ldots
\end{equation}
The indices $(a,b)$ run over all distinct pairs of $(x,y,z)$, thus rendering the expression $SO(3)$ invariant. Since the Hermitian matrix $\mathbf{\Sigma}\cdot \mathbf{\Sigma}$ has positive eigenvalues, and the commutator $i [\mathbf{\Sigma}^a,\mathbf{\Sigma}^b]$ is Hermitean, we see that non-zero commutations cost potential energy. Thus, at least locally the potential \eqref{veffsigma} is minimized when the $\mathbf{\Sigma}$ mutually commute (which means, in turn, that we can mutually diagonalize the $\mathbf{\Sigma}$). In this approximation, we can estimate the minimum value of $V_{eff}$ as the first term in the expansion \eqref{matrixTaylor}. The problem we want to solve becomes a saddle point approximation of the following matrix integral for $M \gg N$:
\begin{equation}
Z[\mathbf{\Sigma}] = \int d\mathbf{\Sigma}^x d\mathbf{\Sigma}^y d\mathbf{\Sigma}^z e^{M \text{tr} \sqrt{\mathbf{\Sigma}\cdot \mathbf{\Sigma} }-  \text{tr} \mathbf{\Sigma}\cdot \mathbf{\Sigma}/4}~.
\end{equation} 
In order to obtain the saddle point equation for the eigenvalues, we first introduce a delta function $\delta(\rho-\mathbf{\Sigma}\cdot \mathbf{\Sigma})$ and integrate out the $\mathbf{\Sigma}$, such that we remain with an integral over the $N\times N$ Hermitian $\rho$ matrix. Upon diagonalizing $\rho$, and including the Vandermonde contribution, we can obtain the potential for its eigenvalues $\rho_i \ge 0$. It is convenient at this point to rescale ${\rho}_i = M^2 \tilde{\rho}_i$. We find:
\begin{equation}
V_{eff}[\tilde{\rho}_i] = -\sum_{j\neq i} \log|\tilde{\rho}_i - \tilde{\rho}_j|  -  M ^2  \sum_i \left(  \sqrt{\tilde{\rho}_i}  -  \frac{\tilde{\rho}_i}{4} + \frac{2 N}{M^2} \log \tilde{\rho}_i  \right) ~,
\end{equation}
up to an additive constant of order $N^2 \log M$. The $\log\tilde{\rho}_i$ contribution comes from the measure of the path integral: there is a Jacobian when changing variables from the $\mathbf{\Sigma}$ matrices to the $\rho$ matrix. The saddle point equation governing the eigenvalues is:
\begin{equation}
\sum_{j \neq i}^N \frac{1}{\tilde{\rho}_i - \tilde{\rho}_j} =  -\frac{2N}{\tilde{\rho}_i} - M^2\left(  \frac{1}{2\sqrt{\tilde{\rho}_i}}  - \frac{1}{4} \right)~.
\end{equation}
To leading order in a large $M$ expansion (taking $M$ to be much larger than $N$) we can consider $\tilde{\rho}_i$ to be peaked around $\tilde{\rho}_i \sim 4$. Expanding about $\tilde{\rho}_i = 4 + \delta_i$ for small $\delta_i$, and keeping the leading term only, we have:
\begin{equation}
\sum_{j \neq i}^N \frac{1}{\delta_i -\delta_j} = \frac{M^2}{32} \, \delta_i~.
\end{equation}
For large\footnote{We are considering here the situation where both $M$ and $N$ are large but $M \gg N$.} $N$, the above eigenvalue equation is solved by the Wigner semicircle distribution \cite{Brezin:1977sv} and has compact support in the interval $(\sqrt{32 N}/M) \times [-1,1]$. Thus, going back to the original eigenvalues, we see that they are peaked around ${\rho}_i \approx 4 M^2$ with a width of order $\sqrt{N} M$. We can approximate the ground state energy to be $V^{(min)}_{eff} \approx - M^2 N$. It would be interesting to study subleading corrections, due to the repulsion of eigenvalues from the Vandermonde, but we will not do so here. 


There is a slightly more efficient way to see the above. Using the property $\text{tr} \, \mathbf R^2 = 2 \, \text{tr} \, \mathbf{\Sigma} \cdot \mathbf{\Sigma}$ we can write the effective potential (\ref{veffsigma}) completely in terms of the eigenvalues of $\mathbf R$ as:
\begin{equation}
V_{eff} = \frac{1}{2} \sum_{n=1}^{2N} \left( - M  |\lambda_n| + \frac{\lambda_n^2}{4} \right)~.
\end{equation}
Again, at least in the limit $M \gg N$ where we can ignore the effects of the matrix measure, we find $V_{eff}^{(min)} \approx -M^2 N$ as before. 

We now proceed to study the kinetic contribution linear in velocity. 

\subsubsection{Linear velocity term}

We consider the linear velocity term for the matrix model. The simplest case occurs when the $\mathbf{\Sigma}_{ij}$ matrix is diagonal, i.e. $\mathbf{\Sigma}_{ij} = \mathbf{x}_i \, \delta_{ij}$ with $i=1,\ldots,N$. In this case, we simply find a sum of $N$ terms (one for each $\mathbf{x}_i$) each identical with the vector case. Each will have their own $M+1$ lowest Landau levels. 
Generally, however, the $\mathbf{\Sigma}^a$ will not be mutually diagonalizable. Inspired by the expression (\ref{blochkin}), 
we claim that the  linear velocity term is given by:
\begin{equation}\label{matrixwz}
S_{kin}^{(1)} =  i \, \frac{M}{2} \, \text{tr} \,  \int dt \, \left[ {\dot{Z}^\dag   \left(  \mathbb{I} + Z {Z}^\dag \right)^{-1}  {{Z}} - {Z}^\dag  \left(  \mathbb{I} + Z {Z}^\dag \right)^{-1}  \dot{Z}   } \right] ~,
\end{equation}
where $Z_{ij}$ is a complex $N\times N$ matrix. The stereographic map \eqref{eq:spinExpectation} relating $z$ to a point on the Bloch sphere is generalized to:
\begin{eqnarray}\label{stereo}
\mathbf{\Sigma}^x + i \mathbf{\Sigma}^y  &\equiv&  2M \,  Z \, (\mathbb{I}+ Z^\dag Z)^{-1}~, \\
\mathbf{\Sigma}^x - i \mathbf{\Sigma}^y   &\equiv&  2M \, Z^\dag \, (\mathbb{I}+ Z Z^\dag)^{-1}~,  \\ \label{stereo3}
\mathbf{\Sigma}^z &\equiv& M \, \left[ \mathbb{I} -  \, \left( \mathbb{I} +  Z Z^\dag \right)^{-1}  - \left( \mathbb{I} +  Z^\dag Z \right)^{-1} \right]~.
\end{eqnarray}
In order to verify that $\mathbf{\Sigma}^a = (\mathbf{\Sigma}^a)^\dag$ it is useful to take advantage of identities such as: $(\mathbb{I}+Z Z^\dag)^{-1} Z = Z (\mathbb{I}+Z^\dag Z)^{-1}$. 
Naturally, when $N=1$ our expression (\ref{matrixwz}) reduces to the expression (\ref{blochkin}). It is also time reparameterization invariant under $\tau \to f(\tau)$ and $Z_{ij}(\tau) \to Z_{ij}(f(\tau))$. Moreover, our expression is invariant under the global $SU(N)$, under which $Z \to \Lambda Z \Lambda^\dag$, with $\Lambda \in SU(N)$. In fact, as we shall see in the next subsection, (\ref{matrixwz}) invariant under a larger group $U(2N)$ acting as:
\begin{equation}\label{u2n}
Z \to (A Z+B)(C Z +D)^{-1}~, \quad\quad \begin{pmatrix}
  A & B \\
  C & D 
 \end{pmatrix} \cdot  \begin{pmatrix}
  A & B \\
  C & D 
 \end{pmatrix}^\dag = \mathbb{I}_{2N \times 2N}~. 
\end{equation}
where $A$, $B$, $C$ and $D$ are $N\times N$ matrices. The $U(2N)$ invariance is in agreement with our observation that terms stemming from the functional determinant (\ref{fdetu2n}) exhibit a $U(2N)$ symmetry. This generalizes the $SU(2)$ symmetry (\ref{ztransform}) that is present in the $N=1$ case. Recall that in the $N=1$ case, the linear velocity term only depended on two of the three variables in $\mathbf{x}$. Analogously, our expression (\ref{matrixwz}) only depends on $2N^2$ of the $3N^2$ variables in the three Hermitian matrices $\mathbf{\Sigma}^a$. 



\subsection{Berezin coherent states}

As in the vector case, the matrix action (\ref{matrixwz}) can stem from a curved phase space endowed with a K{\"a}hler structure. These compact K{\"a}hler manifolds were studied extensively by Berezin \cite{Berezin:1978sn}. The K{\"a}hler metric is given by:
\begin{equation}\label{Kmetric}
ds^2 = M \, \text{tr} \, dZ \left(  \mathbb{I} +Z Z^\dag \right)^{-1} dZ^\dag \left( \mathbb{I} +Z^\dag Z \right)^{-1}~,
\end{equation}
where $c$ is a normalization constant. The K{\"a}hler potential is given by:
\begin{equation}
K = M \, \log\left( \mathbb{I}  + Z Z^\dag \right)~.
\end{equation}
This potential transforms under the $U(2N)$ isometry \eqref{u2n} as
\begin{align}
    K &\rightarrow K - M \log\det( Z^\dag C^\dag + D^\dag ) - M \log\det( CZ + D )    \ ,
\end{align}
leaving the metric (\ref{Kmetric}) invariant. It is the natural generalization of the $N=1$ case. 

More precisely, what Berezin shows \cite{Berezin:1978sn} is that there exist a collection of coherent states, analogous to the Bloch coherent states, parameterized by a complex matrix $Z_{ij}$. Explicitly:
\begin{equation}
|Z^\dag_{ij}\rangle = \frac{e^{Z^\dag_{ij} \mathbf{\hat{S}}^+_{ji}}}{\det( \mathbb{I} +Z^\dag Z)^{M/2}} |v\rangle~, \quad\quad \mathbf{\hat{S}}^\pm_{ij} = \mathbf{\hat{S}}^x_{ij} \pm i\mathbf{\hat{S}}^y_{ij}~,
\end{equation}
where the state $|v\rangle$ is the state annihilated by all $\psi^{1}_{Ai}$ and $\bar{\psi}^{2}_{iA}$ operators. It can be expressed as $|v\rangle = \prod_{A,i} \bar{\psi}_{iA}^2 |0\rangle$, where $|0\rangle$ is the state that is annihilated by all the $\psi_{Ai}^\alpha$ operators. Consequently $|v\rangle$ is annihilated by $\hat{\mathbf S}_{ij}^-$. The overlap between two Berezin coherent states is given by:
\begin{equation}
\langle W_{ij} | Z_{ij}^\dag \rangle = \frac{\det \left( \mathbb{I} +  W Z^\dag \right)^M}{\det\left( \mathbb{I} + W^\dag W\right)^{M/2}\det\left( \mathbb{I} + Z^\dag Z\right)^{M/2}}~.
\end{equation}
At large $M$ the quantum evolution of a certain class of $U(M)$ invariant operators in the Grassmann theory becomes approximately classical with an emergent curved phase space \cite{Berezin:1978sn} , the geometry of which is described by the K{\"a}hler metric (\ref{Kmetric}). The role of large $M$ becomes that of the small Planck constant. The classical Hamiltonian, governing the time evolution of functions on the emergent phase space, is given by $H[Z,Z^\dag] = \langle Z | \hat{H} | Z^\dag\rangle$.
The volume of the emergent classical phase space computes the number of quantum states obtained upon quantizing it. The number of quantum states was computed in \cite{Das:2012dt}. The result reads:
\begin{equation}\label{dimHBerezin}
\dim {\mathcal{H}_K} = \prod_{j=1}^N \frac{\Gamma[N+M+j]\Gamma[j]}{\Gamma[N+j]\Gamma[M+j]}~.
\end{equation} 
We can study the behavior of $\dim {\mathcal{H}_K}$ in various limits. When $N \gg M \gg 1$ we find $\dim {\mathcal{H}_K} \sim 2^{2MN}$ to leading order. Thus in this limit, the dimension of the effective Hilbert space closely approximates the full Hilbert space of the original Grassmann system. For $M \gg N \gg 1$ we find instead $\dim {\mathcal{H}_K} \sim M^{N^2}$. Finally, for $M  = \alpha N$ where $\alpha$ is fixed in the large $N$ limit, we have:
\begin{equation}
\log \dim \mathcal{H}_K = f(\alpha) N^2 + \ldots
\end{equation} 
with:
\begin{equation}
f(\alpha) =  \frac{1}{2} \left(\alpha ^2 \log (\alpha )-2 (\alpha +1)^2 \log (\alpha +1)+(\alpha +2)^2 \log (\alpha +2)-2 \log 4 \right)~.
\end{equation}
Notice that in the limit $\alpha \to 0$, $f(\alpha) \sim 2 \alpha \log2$ for which $\log \dim \mathcal{H}_K \sim 2 N M \log 2 $. Similarly, in the $\alpha \to \infty$ limit, $f(\alpha) \sim \log\alpha$ for which $\log \dim \mathcal{H}_K \sim  N^2 \log M$. As shown in the appendix, (\ref{dimHBerezin}) is precisely the number of states we would obtain in the Grassmann matrix model, had we gauged the $U(M)$ global symmetry. This is to be expected. The full space of $U(M)$ invariant states can be built by acting with a function of the $U(M)$ invariant operator $\mathbf{\hat{S}}^+_{ij}$ on the state $| v \rangle$ (which is itself defined to be $U(M)$ invariant by a suitable choice of the normal ordering constant in the $U(M)$ generators). 




\subsubsection{Hamiltonian and path integral}


In the vector case, the Hamiltonian $\hat{H}$ (\ref{vecham}) we studied was constant along the Bloch two-sphere given that all the Bloch coherent states had the same total angular momentum. In this regard our matrix model differs from the vector case. Given our Hamiltonian operator (\ref{matrixH}), the Hamiltonian $H[Z,Z^\dag] \equiv \langle Z | \hat{H} | Z^\dag  \rangle$ governing time evolution on the emergent classical phase space is found to be:
\begin{equation}
H[Z,Z^\dag] = -N M^2 + M^2 \, \text{tr} \, (S^0)^2~,
\end{equation}
to leading order in $M$. We have defined:
\begin{equation}
S^0 \equiv \left[ \left( \mathbb{I} + Z  Z^\dag \right)^{-1} - \left(  \mathbb{I} + Z^\dag Z \right)^{-1} \right]~.
\end{equation}
Notice that $H[Z,Z^\dag]$ is invariant under $Z \to U Z U^\dag$ where $U \in SU(N)$. Moreover, the Hamiltonian $H[Z,Z^\dag]$ is minimized when $Z$ and $Z^\dag$ commute, where it takes the value $E_{min} = -N M^2$. Consequently, the state $| v\rangle$ is one of these minimal energy states. This agrees with our analysis of the effective potential in section \ref{matrixveff}, where the minimum was also found to be $-N M^2$ in the large $M$ limit. When $Z$ and $Z^\dag$ commute they can be mutually diagonalized and the K{\"a}hler metric becomes $N$ copies of $\mathbb{C}\mathbb{P}^1$, i.e. one Bloch sphere for each eigenvalue. Furthermore, as was found in the analysis of section \ref{matrixveff}, the commutator of $Z$ and $Z^\dag$ costs energy. Nevertheless, since the $Z$ can be continuously deformed, there is a rich low energy sector continuously connected to the ground states given by almost commuting complex matrices.

Given the kinetic term and the Hamiltonian on phase space, following Berezin \cite{Berezin:1978sn}, we can write down the real time path integral for transition amplitudes between coherent states $| Z^\dag_i \rangle$ and $\langle Z_f |$. It reads:
\begin{equation}
\mathcal{A}_{fi} = \int \mathcal{D} \mu[Z,Z^\dag]  \exp \left(\frac{ M}{2} \text{tr} \int_{-T}^T dt  \, \left[ \dot{Z}  (\mathbb{I} +  Z^\dag Z)^{-1} {Z}^\dag - h.c. \right] - i \int_{-T}^T dt  H [Z,Z^\dag] \right)~,
\end{equation}
with boundary conditions $Z^\dag[-T] = Z^\dag_i$ and $Z[T] = Z_f$. The measure factor is given by:
\begin{align}
    \mathcal{D} \mu[Z,Z^\dag] &\equiv \frac{1}{\N} \frac{\mathcal{D} Z \, \mathcal{D}  Z^\dag  }{ \det \left(\mathbb{I}+ Z Z^\dag \right)^{2N}}~. 
\end{align} 
The normalization constant $\N$ ensures that $\Tr\, \mathbb{I} = \int{{d} \mu[Z, Z^\dag] } = \dim \mathcal{H}_K$. It can be computed by use of the Selberg integral $S_N(1, M+1, 1)$ \cite{selberg}. 

Consider finally the following rescaling $Z = {M}^{-1/2} \tilde{Z}$, with $\tilde{Z}$ fixed in the large $M$ limit, and in addition $M\gg N$. To leading order in the large $M$ expansion, the path integral becomes: 
\begin{multline}\label{Zfluct}
\mathcal{A}_{fi} = \int \mathcal{D} \tilde{Z} \mathcal{D} \tilde{Z}^\dag \, \exp \left[ \frac{1}{2} \text{tr} \int_{-T}^T dt \, \left( \dot{\tilde{Z}} {\tilde{Z}}^\dag - h.c. \right)-  i \, \text{tr} \, \int_{-T}^T dt \, [\tilde{Z},\tilde{Z}^\dag]^2 \,  \right]~.
\end{multline}
This limit is a small fluctuation limit in which the geometry of the curved phase space becomes flat and the Hamiltonian boils down to the trace of the square of the commutator. Naturally, in the $N=1$ case, no such commutator arises, and the rescaling limit simply describes motion in a small flat patch of the full $\mathbb{C}\mathbb{P}^1$. 
\newline\newline
Thus, we generalize several of the features observed in the vector model to the matrix model. As before, there is an emergent classical phase space endowed with a K{\"a}hler metric, a low velocity expansion of a bosonic Hermitian matrix model in a suitable large $M$ regime and a large number of low energy states. Given the appearance of a bosonic matrix model, we can wonder about a holographic interpretation at large $N$. We end with some speculative remarks on this question.  



\section{Outlook}

We have discussed systems with a finite dimensional Hilbert space, whose constituents are a large number of spin-1/2 fermions. For certain collections of states, we have seen how the systems we have considered exhibit an emergent classical phase space parameterized by complex coordinates. The phase space is endowed with a K{\"a}hler metric which in the simplest case is nothing more than the round two-sphere. More generally, it is a complex matrix generalization thereof. In the vector case, the size of the Bloch sphere (\ref{blochsphere}) scales as the logarithm of the dimension of the Hilbert space. The specific Hamiltonian we considered, commutes with the total angular momentum operator. Consequently, transition amplitudes between different Bloch coherent states lie on a Bloch sphere of fixed size. One manifestation of this is that the parameter $r$ acquires no time derivatives in the effective action. More generally, one might imagine Hamiltonians with matrix elements connecting Hilbert spaces with different total angular momenta. In such a case, one might consider an additional direction given by the size of the two-sphere, such that in a suitable large $M$ limit, the low energy degrees of freedom are parameterized by coordinates in a three-dimensional ball. So long as the dimension of the Hilbert space remains finite, there is still a cap on the maximal size of the two-sphere. A natural matrix generalization of the parameter $r$ is given by the trace of the Hermitian matrix $\sqrt{\mathbf{\Sigma}\cdot \mathbf{\Sigma}}$. Unlike the vector case, transitions between different values of $\text{tr} \, \sqrt{ \mathbf{\Sigma} \cdot \mathbf{\Sigma}}$ are possible within the space of Berezin coherent states. In other words, the K{\"a}hler metric of the emergent classical phase space does not constrain $\mathbf{\Sigma}\cdot \mathbf{\Sigma}$ (which is a now a function of $Z$ and $Z^\dag$) to take a specific value. 

Holographically, large $N$ matrix models might be associated with a gravitational theory. For the quantum mechanical model \cite{Itzhaki:1998dd} dual to the ten-dimensional geometry near a collection of $N$ D0-branes, one has nine $N\times N$ Hermitian bosonic matrices $\bold{X}^I_{ij}$ and their Fermionic superpartners. The index $I$ is an $SO(9)$ index, corresponding to the rotational symmetry of the eight-sphere in the near horizon of a stack of $N$ D0-branes in type IIA string theory. The indices $i$ and $j$ run from $1$ to $N$. The Hilbert space is infinite dimensional and there are states with indefinitely high energy. In these models, the emergent radial direction has been argued to be captured by the energy scale. At high energies, the quantum mechanics is weakly coupled. One manifestation of this, from the bulk viewpoint, is that the size (in the string frame) of the eight-sphere shrinks indefinitely at large radial distances, eventually leading to a stringy geometry. 

Consider now a system where the spectrum is capped, as occurs in the deep infrared of a CFT living on a spatial sphere (due to the curvature coupling of the fields). In such a situation we expect the emergent sphere to cap off. This is indeed what happens in global anti-de Sitter space where the sphere at fixed $r$ and $t$ smoothly caps off in the deep interior.\footnote{Recall the metric of global AdS$_{d+2}$ is given by $ds^2 = -dt^2 (1+r^2) + dr^2 (1+r^2)^{-1} + r^2 d\Omega_{d}^2$. As $r \to 0$ the $d$-sphere caps off smoothly.} Consider now the geometry of the static patch of four-dimensional de Sitter space:
\begin{equation}
ds^2 = -dt^2(1 - r^2) + \frac{dr^2}{(1 - r^2)} + r^2 d\Omega_2^2~.
\end{equation}
Notice that the size of the two-sphere resides on a finite interval. It smoothly caps off at $r=0$ and is largest at $r=1$ where the cosmological horizon resides. If, somehow, $r$ was an emergent holographic direction related to the energy scale \cite{Anninos:2011af}, then it would seem we have to cap the spectrum both in the infrared as well as the ultraviolet. This would indicate a holographic quantum mechanical dual with a finite number of states \cite{Banks:2006rx,Parikh:2004wh,Dong:2010pm,Li:2001ky,Volovich:2001rt,Heckman:2011qu}, so long as the spectrum is discrete. If moreover we require the holographic model to have a matrix-quantum mechanical sector described by ordinary bosonic matrices, perhaps the systems we have considered above are natural candidates. We postpone the examination of this proposal and the relation to other approaches of de Sitter holography (for an overview see \cite{Anninos:2012qw}) to future work.


\section*{Acknowledgements}

It is a pleasure to thank Tom Banks, Chris Beem, Umut Gursoy, Sean Hartnoll, Juan Maldacena, Nati Seiberg, Douglas Stanford, Herman Verlinde, and especially Diego Hofman and Erik Verlinde for useful discussions. D.A. would also like to thank the Crete Center for Theoretical Physics for its kind hospitality. F.D. is supported in part by the U.S. Department of Energy (DOE) under DOE grant DE-SC0011941. D.A. acknowledges funding from the NSF. R.M. is supported by FWO-Flanders and would like to thank ISCAP and Columbia University for their hospitality.

\appendix

\section{Counting \texorpdfstring{$U(M)$}{U(M)} gauge invariant states}

In this appendix we present the derivation of the formula for the dimension of the Hilbert space of two complex Grassmann matrices $\chi^i_A$ and $\theta_A^i$ with indices ranging from $i =1,\ldots,N$ and $A=1,\ldots,M$. 


Therefore we consider the action:
\begin{equation}
S = \int dt \left[ \bar{\chi}_A^i i \mathcal{D}_t \chi_B^i + \bar{\theta}_A^i i \bar{\mathcal{D}}_t \theta^i_B -  \left(m_1 \bar{\chi}_A^i  \chi_A^i +  m_2 \bar{\theta}_A^i  \theta_A^i \right) \right]~,
\end{equation}
with $\mathcal{D}_t = \partial_t + i A_t$ and $\bar{\mathcal{D}}_t = \partial_t - i A_t$. The gauge field $A_t =  A_t^\delta T^\delta$ is a Hermitian $M\times M$ matrix, with $T^\delta$ the $M^2$ generators of $U(M)$. The Grassmann matrices transform in the (anti-)fundamental representation of $U(M)$ (we pick $\chi_A$, $\bar{\theta}_A$ in the fundamental). We consider the case with $m_1>0$ and $m_2 >0$.  From the Poisson brackets originating from the above action we obtain the anti-commutation relations of fermionic creation/annihilation operators:
\begin{equation}
\{ {\chi}_A^i , \bar{\chi}_B^j \} = \delta_{AB}\delta_{ij}~, \quad\quad \{ {\theta}_A^i , \bar{\theta}_B^j \} = \delta_{AB}\delta_{ij}  \ .
\end{equation}

Integrating out the gauge field gives us $M^2$ constraints:
\begin{equation}
\delta A_t : \quad \bar{\chi}_A T_{AB}^\delta \, \chi_B - \bar{\theta}_A T_{AB}^\delta \, \theta_B  = 0~, \quad \forall \,\, \delta = 1,2,\ldots,M^2
\end{equation}
We define the vacuum state $|0\rangle$ of the theory to be annihilated by all $\chi$ and $\theta$ operators. Note that it obeys the gauge constraint and is thus gauge invariant. Moreover, acting with gauge invariant operators always increases the energy, hence $|0\rangle$ is unique.

We wish to find the thermal partition function and extract the entropy $S(T)$ at infinite temperature. We can then use the fact that $\lim_{T\to \infty} S(T) = \log \, \dim \mathcal{H}$ to find the dimension of the Hilbert space with a $U(M)$ singlet constraint imposed. In the absence of the gauge field $A_t$, we would have $\dim \mathcal{H} = 2^{2 N M}$.


\subsection{Euclidean path integral}

We can compute the thermal partition function as a Euclidean path integral. Wick rotate time $t \to -i \tau$ such that
\begin{equation}
S_E = \int_0^\beta d\tau \left[ \bar{\chi}_A^i \mathcal{D}_\tau \chi_B^i + \bar{\theta}_A^i  \bar{\mathcal{D}}_\tau \theta^i_B +  m_1 \bar{\chi}_A^i  \chi_A^i +  m_2 \bar{\theta}_A^i  \theta_A^i\right]~.
\end{equation}
The Grassmann variables obey anti-periodic  boundary conditions around the thermal circle. The Euclidean path integral of interest is:
\begin{equation}
Z[\beta] = \int \mathcal{D}A_\tau \mathcal{D} \chi \mathcal{D} \bar{\chi}  \mathcal{D} \theta \mathcal{D} \bar{\theta} \, e^{-S_E}~.
\end{equation}
The gauge transformations acting on $A_\tau$ are given by $A_\tau \to U A_\tau U^\dag + i\partial_\tau U \cdot U^\dag$. Due to the non-contractible thermal circle, we can only fix the gauge up to the holonomy around the thermal circle \cite{Aharony:2003sx}. The Fadeev-Popov procedure in doing so gives us the following action for the (time independent upon gauge fixing) eigenvalues of $A_\tau$ which we denote $\alpha_A$:
\begin{equation}
\int  \prod^M_{A=1}  d \alpha_A  \, \left( \prod_{A<B} \sin^2  \frac{\beta(\alpha_A-\alpha_B)}{2} \right) ~.
\end{equation}
We have dropped an overall constant which we must later recover by computing the zero temperature entropy, which should vanish because the ground state is unique. We have yet to calculate the contribution to the action of the fundamental matter fields. We first expand them in a Fourier expansion:
\begin{align}
    \chi(\tau) &= \sum_{n\in\mathbb{Z}} e^{i 2\pi (n+1/2)\tau/\beta} \chi_n  \ , & \theta(\tau) &= \sum_{n\in\mathbb{Z}} e^{i 2\pi (n+1/2)\tau/\beta} \theta_n    \ .
\end{align}
Thus we obtain the thermal eigenvalues:
\begin{align}
    \lambda^{A}_n &= 2\pi (n+1/2)/\beta + i m_1 + \alpha_A   \ , &   \tilde{\lambda}^{A}_n &= 2\pi (n+1/2)/\beta + i m_2 - \alpha_A   \ .
\end{align}
The determinant to be evaluated is given by $\prod_n \lambda^{A}_n \tilde{\lambda}^{A}_n$. It is UV divergent. We regulate the logarithm of the determinant 
by taking two derivatives with respect to $m$ and integrating $m$ twice while setting the integration constants to zero. The result is:
\begin{equation}
    \sum_n \log \lambda^A_n \tilde{\lambda}^A_n = \log \cos\left(\frac{\beta( i m_1 + \alpha_A)}{2}\right) + \log \cos\left(\frac{\beta( i m_2 - \alpha_A)}{2}\right)~.
\end{equation}
Our remaining integral becomes (we are rescaling the eigenvalues by a factor of the temperature in obtaining the below formula):
\begin{multline}
    Z[\beta] = \mathcal{N} \int \prod_A \,  d\alpha_A \, \prod_{A<B} \sin^2 \left(\frac{\alpha_A-\alpha_B}{2}\right) \\
    \times { \prod_A  \cos^N \left(\frac{i \beta m_1+ \alpha_A}{2}\right)} \, { \cos^N \left(\frac{i \beta  m_2 - \alpha_A}{2}\right)}~.
\end{multline}
Our task has been reduced to solving a multi-variable integral for the $N$ variables $\alpha_A$. To compute the constant $\mathcal{N}$ we fix the ground state to have vanishing energy and due to its uniqueness, we have: $\lim_{\beta \to \infty} Z[\beta] = 1$. The integrals required were solved by Selberg \cite{selberg}. For instance we have for the $\beta =0$ integral (see (1.17) of \cite{selberg}):
\begin{multline}
\int_{-\pi}^\pi \left( \prod_{A=1}^M \frac{d\alpha_i}{2\pi} \right) \, \prod_{A<B} \sin^2 \frac{(\alpha_A- \alpha_B)}{2} \,  \prod_A \cos^{2N} \frac{\alpha_A}{2} = \\ 
2^{-2MN -M(M-1)} \, \prod_{j=1}^M \frac{\Gamma[2N+j]\Gamma[1+j]}{\Gamma[N+j]\Gamma[N+j]}~.
\end{multline}
We can use the same formula with $N=0$ to fix the normalization by considering the $\beta \to \infty$ limit. Thus, using the Selberg integrals, we obtain the final result:
\begin{equation}
\dim \mathcal{H} = \frac{1}{\Gamma[M+1]} \, \prod_{j=1}^M \frac{\Gamma[2N+j]\Gamma[1+j]}{\Gamma[N+j]\Gamma[N+j]}~. 
\end{equation}
Some algebraic manipulations show that the above expression is in fact equivalent to (\ref{dimHBerezin}) as can be easily checked numerically for several cases. Some simple checks are also possible. For $N=1$ we find $\dim \mathcal{H} = (M+1)$. These states are given by acting with powers of $\bar{\chi}_A\bar{\theta}_A$ on $|0\rangle$.

\section{Modified vector model}\label{modvec}

In this appendix we briefly mention a slight modification of the vector model considered in the main body of the text. The degrees of freedom are given by two sets of $M$ complex fermion spinors $\{\psi^\alpha_A, \theta^\alpha_A \}$. We consider the following Euclidean action:
\begin{equation}
S_E = \int d\tau \, \bar{\psi}_A^\alpha \partial_\tau \psi^\alpha_A + \bar{\theta}_A^\alpha \partial_\tau \theta^\alpha_A - \left( \bar{\psi}^\alpha_A \sigma_{\alpha\beta} \psi^\beta_A -  \bar{\theta}^\alpha_A \sigma_{\alpha\beta} \theta^\beta_A \right)^2~.
\end{equation}
Following the procedure outlined in the main text, we end up with an effective action for a bosonic three-vector $\mathbf{x}$:
\begin{equation}\label{seffmod}
S_{eff} = M \text{Tr} \log \left( -\partial_\tau + \mathbf{x} \cdot \sigma \right) + M \text{Tr} \log \left( -\partial_\tau - \mathbf{x} \cdot \sigma \right) + \frac{1}{4} \,  \int d\tau \, r^2~.
\end{equation}
Performing a small velocity expansion one realizes that the term linear in velocity in fact cancels. This is due to the relative sign in front of $\mathbf{x}$ in the two functional determinants in (\ref{seffmod}). Thus the leading term in the velocity expansion is:
\begin{equation}\label{seffmodkin}
S_{kin}^{(2)} = M \int d\tau \frac{1}{4r} {\left( \dot{\theta}^2 + \sin^2 \theta  \, \dot{\phi}^2 \right)}~.
\end{equation}
The reason for the cancellation is that this model has a Hamiltonian given by the difference in angular momentum. The ground state is given by the configuration where the two angular momenta, whose operators are given by $\hat{\mathbf{J}}_1 = \bar{\psi}_A \sigma \psi_A/2$ and $\hat{\mathbf{J}}_2 = \bar{\theta}_A \sigma \theta_A/2$, are anti-aligned. In the language of the charged particle on the two-sphere, it is as if we have added a positron on top of the electron, thus canceling the effect of the Lorentz force, leaving an ordinary kinetic term for the bound neutral particle. The configuration space is still parameterized by the angles on a two-sphere. The mass of the neutral particle is twice that of the original one, explaining the $1/4$ as opposed to the $1/8$ in (\ref{seffmodkin}). As before, at large $M$ we have a controlled low velocity expansion. At high energies, the two angular momenta can fluctuate independently and this simple picture is lost. A similar modification can be made for the matrix model. 


\end{document}